\documentclass[prl,floatfix,twocolumn,showpacs,preprintnumbers,amsmath,amssymb,superscriptaddress]{revtex4}
\usepackage{graphicx,float}
\usepackage[ansinew]{inputenc}

\newcommand{\udt}[3]{#1^{#2}_{\phantom{#2}#3}}
\newcommand{\dut}[3]{#1_{#2}^{\phantom{#2}#3}}

\begin{document}

\thispagestyle{empty}

\begin{center}
\title{\large{\bf Rotating Charged Cylindrical Black Holes as Particle Accelerators}}
\date{20th April 2011}
\author{Jackson Levi Said\footnote{jacksons.levi@gmail.com}}
\affiliation{Physics Department, University of Malta, Msida, MSD 2080, Malta}
\author{Kristian Zarb Adami\footnote{kris.za@gmail.com}}
\affiliation{Physics Department, University of Malta, Msida, MSD 2080, Malta}
\affiliation{Physics Department, University of Oxford, Oxford, OX1 3RH, United Kingdom}

\begin{abstract}
{It has recently been pointed out that arbitrary center-of-mass energies may be obtained for particle collisions near the horizon of an extremal Kerr black hole. We investigate this mechanism in cylindrical topology. In particular we consider the center-of-mass energies of a cylindrical black hole with an extremal rotation and charge parameter. The geodesics are first derived with a rotating charged cylindrical black hole producing the background gravitational field. Finally the center-of-mass is determined for this background and its extremal limit is taken.}
\end{abstract}


\maketitle

\end{center}

\section{I. Introduction}
\begin{figure}
\centerline{\includegraphics[width=4cm,height=6cm]{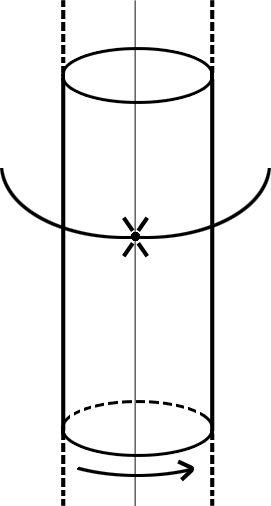}}
\caption{The experimental set up is shown. A rotating charged cylindrical black hole is considered and a pair of neutral particles allowed to collide at various radii away from the event horizon.}
\label{cyl_top}
\end{figure}

It was recently suggested by Bañados, Silk, and West (BSW) in Ref.\cite{p4} that a rotating spherical black hole acts like a particle accelerator in the center-of-mass frame of the collision of a pair of particles. In particular BSW found that as the rotation parameter of the black hole in question becomes extremal and the collision moves onto the horizon, the energy tends to arbitrarily high values. This could be one of the only ways in which Planck scale physics could be probed seeing as no current particle accelerator design can explore this scale of physics. However in the case of Ref.\cite{p4} the particles must have very specific angular momenta in order to achieve this result. On the other hand Ref.\cite{p5,p6} point out that astrophysical black holes contain within them deviations of the extremal rotation parameter first pointed out by Ref.\cite{p7}. Thus arbitrary center-of-mass energies may not truly be realized in astrophysical black holes, but this does not leave out the possibility that mini-black holes could reach the extremal rotation parameter.
\newline
The universal property of acceleration of particles was investigated in Ref.\cite{p8} for pairs of particles. The BSW mechanism was also generalized for charged Kerr, or Kerr-Newman, black holes in Ref.\cite{p9} with neutral particles giving the same result as in Ref.\cite{p4}, as was expected, however similar limitation were found when the rotation parameter limit of Ref.\cite{p7} was considered. Lastly the case of nonrotating black holes with charged particles was investigated in Ref.\cite{p10}, where the same mechanism was observed for even the simplest case of just radial motion.
\newline
In this paper we investigate the BSW mechanism in a wholly different type of black hole, one with a different topology. In Ref.\cite{p4} the $\mathcal{S}^2$ topology was considered, we will explore the $\mathcal{R}\times\mathcal{S}^1$ topology, namely the cylindrical black hole, as shown in Fig.(\ref{cyl_top}). The anti-de Sitter rotating cylindrical black hole metric from Ref.\cite{p1} is used in the derivation. Repeated indices are to be summed and natural units, $G=1=c$ are assumed throughout. Lastly the signature $\left(-,+,+,+\right)$ is taken. As in all the above referred to papers we neglect the effects of gravitational waves and back reaction.

\section{II. Cylindrical Black Holes}
\subsection{The Geodesic Equations}
The rotating charged cylindrical black hole is derived by considering the Einstein-Hilbert action in four dimensions with a cosmological constant. By inserting a cylindrically symmetric generic metric into this action, the metric components are derived. The explicit metric is determined in Ref.\cite{p1}, with related work to be found in Ref.\cite{p11,p12}, which found it to be given by
\begin{eqnarray}
ds^2& &=-\Delta\left(\gamma dt-\frac{\omega}{\alpha^2}d\phi\right)^2+r^2\left(\omega dt-\gamma d\phi\right)^2\nonumber\\
& &+\frac{dr^2}{\Delta}+\alpha^2r^2dz^2
\label{metric}
\end{eqnarray}
where
\begin{eqnarray}
\Delta&=&\alpha^2r^2-\frac{b}{\alpha r}+\frac{c^2}{\alpha^2r^2}\\
b&=&4M\left(1-\frac{3}{2}a^2\alpha^2\right)\\
c^2&=&4Q^2\left(\frac{1-\frac{3}{2}a^2\alpha^2}{1-\frac{1}{2}a^2\alpha^2}\right)\\
\gamma&=&\sqrt{\frac{1-\frac{1}{2}a^2\alpha^2}{1-\frac{3}{2}a^2\alpha^2}}\\
\omega&=&\frac{a\alpha^2}{\sqrt{1-\frac{3}{2}a^2\alpha^2}}
\end{eqnarray}
with
\begin{eqnarray}
-\infty<&z&<+\infty\nonumber\\
0\leq&r&<+\infty\nonumber\\
0\leq&\phi&<2\pi
\end{eqnarray}
The cylindrical topology is spanned by $\left(r,\phi\right)$ for the polar part and $\left(z\right)$ for the axial part. Now since anti-de Sitter spacetime is being considered, $\alpha^2=-\frac{1}{3}\Lambda$ results in a real $\alpha$. Lastly the rotation parameter, $a$, can range between the following
\begin{equation}
0\leq a\alpha\leq1
\end{equation}
The inverse of this metric is given by
\begin{eqnarray}
\frac{\partial^2}{\partial s^2}&=&-\frac{r^2\gamma^2\alpha^4-\Delta\omega^2}{r^2\Delta\left(\omega^2-\alpha^2\gamma^2\right)^2}\frac{\partial^2}{\partial t^2}\nonumber\\
& &-2\frac{\alpha^2\gamma\left(r^2\alpha^2-\Delta\right)\omega}{r^2\Delta\left(\omega^2-\alpha^2\gamma^2\right)^2}\frac{\partial}{\partial t}\frac{\partial}{\partial\phi}+\Delta\frac{\partial^2}{\partial r^2}\nonumber\\
& &+\frac{\alpha^4\left(\gamma^2\Delta-r^2\omega^2\right)}{r^2\Delta\left(\omega^2-\alpha^2\gamma^2\right)^2}\frac{\partial^2}{\partial\phi^2}+\frac{1}{\alpha^2r^2}\frac{\partial^2}{\partial z^2}\nonumber\\
& &
\label{imetric}
\end{eqnarray}
This will be useful when performing calculations.
\newline
Now following \cite{p2,p3,b1}, we take the Lagrangian
\begin{equation}
\mathcal{L}=\frac{1}{2}g_{\mu\nu}\dot{x}^{\mu}\dot{x}^{\nu}
\end{equation}
where dots denote covariant differentiation with respect to an affine parameter $\lambda$. One imposes the condition
\begin{equation}
\tau=m_0\lambda
\end{equation}
where $\tau$ is the proper time as measured by the particle and $m_0$ is the mass of a particle in the gravitational field. This is equivalent to imposing the normalization condition
\begin{equation}
g_{\mu\nu}\dot{x}^{\mu}\dot{x}^{\nu}=-\dut{m}{0}{2}
\end{equation}
For zero and negative values of $\dut{m}{0}{2}$, the Lagrangian can be used to derive geodesics for null and spacelike geodesics, however timelike geodesics will be considered in this paper. In order to consider the Hamiltonian formularism the following momenta are taken
\begin{equation}
p_{\mu}=\frac{\partial\mathcal{L}}{\partial\dot{x}^{\mu}}=g_{\mu\nu}\dot{x}^{\nu}
\end{equation}
which gives the Hamiltonian
\begin{eqnarray}
H&&=-\mathcal{L}=-\frac{1}{2}g_{\mu\nu}\dot{x}^{\mu}\dot{x}^{\nu}\nonumber\\
&&=\frac{1}{2}g^{\mu\nu}\left(g_{\mu\sigma}\dot{x}^{\sigma}\right)\left(g_{\lambda\nu}\dot{x}^{\lambda}\right)\nonumber\\
&&=\frac{1}{2}g^{\mu\nu}p_{\mu}p_{\nu}
\end{eqnarray}
By associating the momentum with the first derivative of the Hamilton-Jacobi action, $S$, with respective to the corresponding coordinate, i.e. $p_{\left(x^{\mu}\right)}=\partial S/\partial x^{\mu}$, the Hamilton-Jacobi equation for geodesics can be calculated. Starting with the Hamilton-Jacobi equation
\begin{equation}
-\frac{\partial S}{\partial\lambda}=H=\frac{1}{2}g^{\mu\nu}\frac{\partial S}{\partial x^{\mu}}\frac{\partial S}{\partial x^{\mu}}
\label{hamjac}
\end{equation}
From the symmetries of the background spacetime of a cylindrical black hole, there are two constants of motion that must be preserved for any geodesic, namely the energy, $E$, of each particle, and the angular momentum, $L$, of each particle, which can be related to the four-momentum as follows
\begin{eqnarray}
p_t&=&-E\\
p_{\phi}&=&L
\end{eqnarray}
Furthermore since we are considering cylindrical geometry, the axial component is taken as null. This does not reduce the generality of the result because the background spacetime is not curved by the black hole in this coordinate and for any nonzero initial point, the black hole coordinate system can be transformed to zero by an appropriate transformation.
\newline
Thus if a separable solution exists for the Hamilton-Jacobi equation then it must be of the form
\begin{equation}
S=\frac{1}{2}\dut{m}{0}{2}\lambda-Et+L\phi+S_r\left(r\right)
\label{jaction}
\end{equation}
Now using the inverse metric in Eq.(\ref{imetric}), the separable form of the Jacobi action in Eq.(\ref{jaction}) and the Hamilton-Jacobi equation in Eq.(\ref{hamjac}), the action turns out to be given by
\begin{equation}
S=\frac{1}{2}\dut{m}{0}{2}\lambda-Et+L\phi+\displaystyle\int^r du R\left(u\right)
\label{jactions}
\end{equation}
where
\begin{eqnarray}
R\left(r\right)&=&\frac{1}{r\Delta\left(\omega^2-\alpha^2\gamma^2\right)}\Big[\left(r^2\gamma^2\alpha^4-\Delta\omega^2\right)E^2\nonumber\\
& &-2\alpha^2\gamma\left(r^2\alpha^2-\Delta\right)\omega E L\nonumber\\
& &-\alpha^4\left(\gamma^2\Delta-r^2\omega^2\right)L^2\nonumber\\
& &-\dut{m}{0}{2}r^2\Delta\left(\omega^2-\alpha^2\gamma^2\right)^2\Big]^{1/2}
\end{eqnarray}
Then differentiating Eq.(\ref{jactions}) with respect to $\dut{m}{0}{2}$, $E$ and $L$, we obtained in integral form the equations of motion, namely,
\begin{eqnarray}
\lambda&&=\displaystyle\int dr\frac{1}{R\left(r\right)\,\Delta}\\
t&&=\nonumber\\
&&\displaystyle\int dr\frac{\left(r^2\gamma^2\alpha^4-\Delta\omega^2\right)E-\alpha^2\gamma\left(r^2\alpha^2-\Delta\right)\omega L}{R\left(r\right)\Delta^2r^2\left(\omega^2-\alpha^2\gamma^2\right)^2}\nonumber\\
&& \\
\phi&&=\nonumber\\
&&\displaystyle\int dr\frac{\alpha^2\gamma\left(r^2\alpha^2-\Delta\right)\omega E+\alpha^4\left(\gamma^2\Delta-r^2\omega^2\right)L}{R\left(r\right)\Delta^2r^2\left(\omega^2-\alpha^2\gamma^2\right)^2}\nonumber\\
&&
\end{eqnarray}
These are more conveniently expressed in differential form as
\begin{eqnarray}
\dot{t}& &=\frac{1}{r^2\Delta\left(\omega^2-\alpha^2\gamma^2\right)^2}\nonumber\\
& &\left[\left(r^2\gamma^2\alpha^4-\Delta\omega^2\right)E-\alpha^2\gamma\left(r^2\alpha^2-\Delta\right)\omega L\right]\label{vel_t}\nonumber\\
& &\\
\dot{\phi}& &=\frac{1}{r^2\Delta\left(\omega^2-\alpha^2\gamma^2\right)^2}\nonumber\\
& &\left[\alpha^2\gamma\left(r^2\alpha^2-\Delta\right)\omega E+\alpha^4\left(\gamma^2\Delta-r^2\omega^2\right)L\right]\label{vel_r}\nonumber\\
& &\\
\dot{r}& &=\frac{1}{r\left(\omega^2-\alpha^2\gamma^2\right)}\nonumber\\
& &\Bigg[\left(r^2\gamma^2\alpha^4-\Delta\omega^2\right)E^2-2\alpha^2\gamma\left(r^2\alpha^2-\Delta\right)\omega E L\nonumber\\
& &-\alpha^4\left(\gamma^2\Delta-r^2\omega^2\right)L^2\nonumber\\
& &-\dut{m}{0}{2}r^2\Delta\left(\omega^2-\alpha^2\gamma^2\right)^2\Bigg]^{1/2}\label{vel_phi}
\end{eqnarray}

\subsection{The Cylindrical Particle Accelerator}
\begin{figure}
\centerline{\includegraphics[width=8cm,height=5cm]{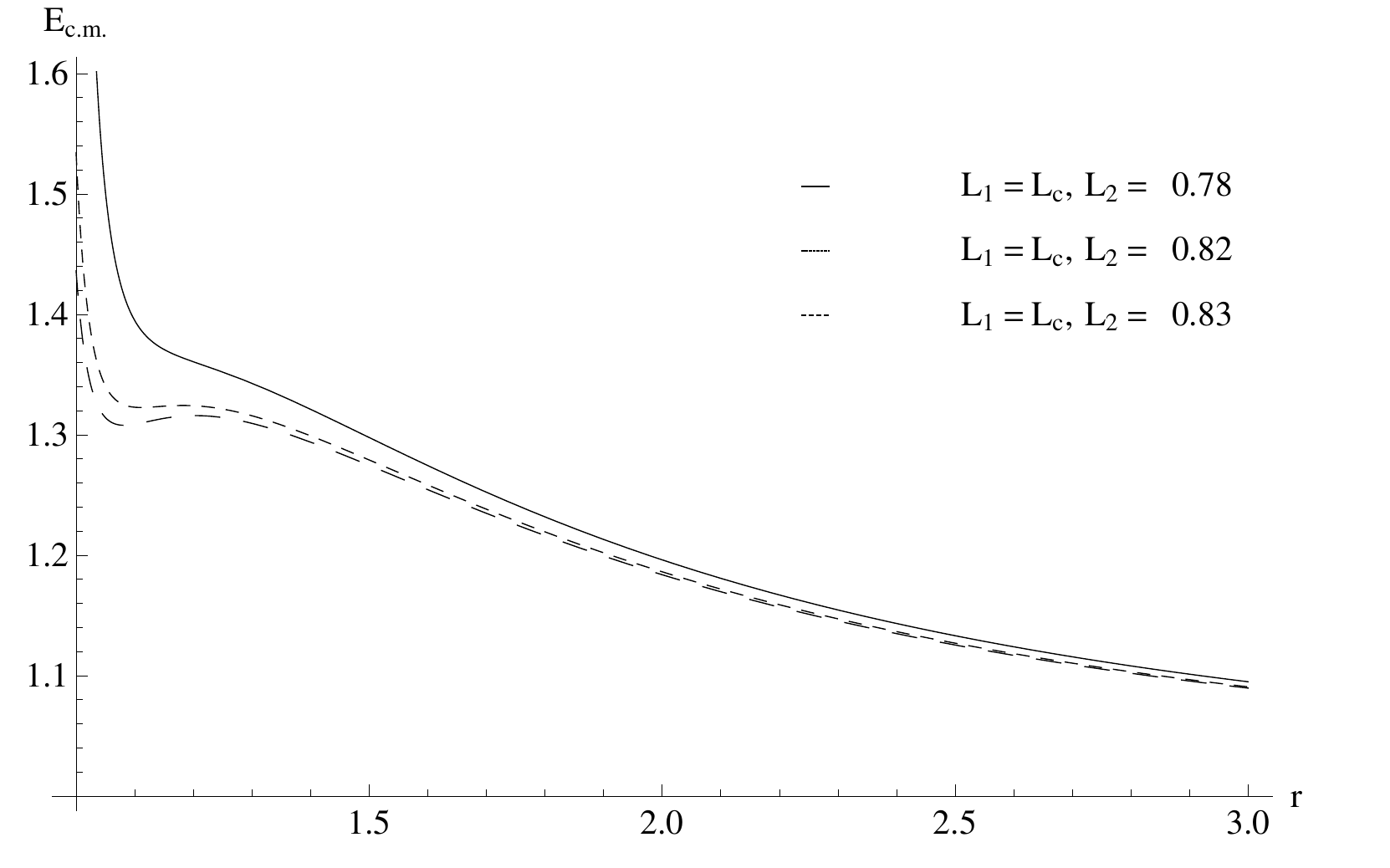}}
\caption{For a rotating cylindrical black hole with extremal rotation and charge parameters (in this case $a=\frac{1}{\sqrt{3}}$ and $Q=\sqrt{\frac{5}{2^{5/3}}}$), the center-of-mass energy is shown against radius all the way up to the horizon for various angular momenta in the two particle collision. The black hole mass is taken to be unity, and $\alpha=\frac{11}{10}$.}
\label{ecm}
\end{figure}

Considering the center-of-mass frame and a pair of particles with an associated mass parameter of $m_0$ and a four-velocity represented by $u_{\left(m\right)}=\left(\dut{u}{{\left(m\right)}}{\mu}\right)$, the collisional energy will be given by
\begin{eqnarray}
E_{c.m.}& &=\left(m_0 u_{\left(1\right)}+m_0 u_{\left(2\right)}\right)^2\nonumber\\
& &=2\dut{m}{0}{2}\left(1+u_{\left(1\right)}\cdot u_{\left(2\right)}\right)\nonumber\\
& &=2\dut{m}{0}{2}\left(1-\eta_{ab}\dut{u}{{\left(1\right)}}{a}\cdot \dut{u}{{\left(2\right)}}{b}\right)
\label{prt1}
\end{eqnarray}
where the normalization condition, $u_{\mu}u^{\mu}=-1$ has been used. Introducing a tetrad basis $\dut{e}{a}{\mu}\left(x\right)$, where Latin indices refer to inertial coordinates and Greek indices to the coordinates in the general coordinate system. The argument in the tetrad refers to the spacetime coordinates under consideration and will be suppressed in the further analysis. The choice of the tetrad frame is constrained by
\begin{eqnarray}
g_{\mu\nu}\dut{e}{a}{\mu}\dut{e}{b}{\nu}=\eta_{ab} & & \udt{e}{a}{\mu}\dut{e}{b}{\mu}=\udt{\delta}{a}{b}
\end{eqnarray}
and must reproduce the general metric by
\begin{equation}
g_{\mu\nu}=\eta_{ab}\udt{e}{a}{\mu}\udt{e}{b}{\nu}
\end{equation}
Lastly the tetrad is ultimately used to transform between vanishing local frames, i.e. Lorentz frames, and general noninertial frames. This is achieved by means of $X^a=\udt{e}{a}{\mu}X^{\mu}$ and $X^{\mu}=\dut{e}{a}{\mu}X^{a}$.
\newline
Now applying the equivalence principle to Eq.(\ref{prt1})
\begin{eqnarray}
\dut{E}{{c.m.}}{2}& &=2\dut{m}{0}{2}\left(1-\eta_{ab}\dut{u}{\left(1\right)}{a}\dut{u}{\left(2\right)}{b}\right)\nonumber\\
& &=2\dut{m}{0}{2}\left(1-g_{\mu\nu}\dut{e}{a}{\mu}\dut{e}{b}{\nu}\dut{u}{\left(1\right)}{a}\dut{u}{\left(2\right)}{b}\right)\nonumber\\
& &=2\dut{m}{0}{2}\left(1-g_{\mu\nu}\left(\dut{e}{a}{\mu}\dut{u}{\left(1\right)}{a}\right)\left(\dut{e}{b}{\nu}\dut{u}{\left(2\right)}{b}\right)\right)\nonumber\\
& &=2\dut{m}{0}{2}\left(1-g_{\mu\nu}\dut{u}{\left(1\right)}{\mu}\dut{u}{\left(2\right)}{\nu}\right)
\label{energy}
\end{eqnarray}
The cylindrical black hole in Eq.(\ref{metric}) will have a corresponding horizon when $1/g_{rr}=0$ which will give horizons at
\begin{equation}
\Delta=0
\label{hor}
\end{equation}
The two particles involved in the collision will have angular momenta $L_1$ and $L_2$, and energies $E_1/m_0=1=E_2/m_0$, for simplicity. Note that the $E_{c.m.}$ is invariant under the transformation $L_1\longleftrightarrow L_2$
\newline
By combining Eq.(\ref{metric}), Eq.(\ref{vel_t}), Eq.(\ref{vel_r}) and Eq.(\ref{vel_phi}) into Eq.(\ref{energy}) gives
\begin{equation}
\frac{\left(E_{c.m.}\right)^2}{2\dut{m}{0}{2}}=1+\frac{A}{3r^2\alpha^2\Delta}
\label{ecm1}
\end{equation}
where
\begin{widetext}
\begin{eqnarray}
A&=&r^2\gamma^2\alpha^4-\Delta\omega^2-\alpha^2\gamma\left(r^2\alpha^2-\Delta\right)\omega\left(L_1+L_2\right)-\alpha^4\left(\gamma^2\Delta-r^2\omega^2\right)L_1L_2-\nonumber\\
& &\sqrt{r^2\gamma^2\alpha^4-\Delta\omega^2-2\alpha^2\gamma\left(r^2\alpha^2-\Delta\right)\omega L_1-\alpha^4\left(\gamma^2\Delta-r^2\omega^2\right)\dut{L}{1}{2}-r^2\Delta\left(\omega^2-\alpha^2\gamma^2\right)^2}\nonumber\\
& &\times\sqrt{r^2\gamma^2\alpha^4-\Delta\omega^2-2\alpha^2\gamma\left(r^2\alpha^2-\Delta\right)\omega L_2-\alpha^4\left(\gamma^2\Delta-r^2\omega^2\right)\dut{L}{2}{2}-r^2\Delta\left(\omega^2-\alpha^2\gamma^2\right)^2}
\end{eqnarray}
\end{widetext}
In the extremal case where both horizons merge, the rotation and charge parameters are constrained by \cite{p1}
\begin{equation}
a^2\alpha^2=\frac{2}{3}-\frac{128}{81}\frac{Q^6}{M^4\left(1-\frac{1}{2}a^2\alpha^2\right)^3}
\end{equation}
For a unit mass black hole with this condition, the event horizon is now given specifically by
\begin{equation}
\alpha r=\frac{1}{\sqrt[3]{2}}
\end{equation}
The physical singularity, on the other hand, is given by
\begin{equation}
\alpha r=0
\label{sing}
\end{equation}
This implies that $r=0$ describes the horizon since $\alpha$ is a nonvanishing constant. Now considering the coordinate transformation
\begin{align}
x&=&r\cos\phi-a\left(1-\frac{a^2\alpha^2}{2}\right)^{-1/2}\sin\phi\\
y&=&r\sin\phi+a\left(1-\frac{a^2\alpha^2}{2}\right)^{-1/2}\cos\phi
\end{align}
which represents a Kerr-Schild-like coordinate transformation \cite{p1} leads to
\begin{equation}
x^2+y^2=\frac{J^2}{M^2-\frac{J^2\alpha^2}{2}}+r^2
\end{equation}
This shows that $r=0$ has internal structure, namely that of a ring which spans the whole $z$ axis which is distinct from the apparent pointlike singularity given in Eq.(\ref{sing}).
\newline
Now the extremal collisional energy appears to diverge for the extremal parameters, however as in Ref.\cite{p4} this is not a true singularity since the numerator also vanishes. Thus applying l'Hôpital's rule, the actual extremal collisional energy is given by
\begin{equation}
\frac{\left(E_{c.m.}\right)^2}{2\dut{m}{0}{2}}=1+\frac{A'}{2r\Delta\left(\omega^2-\alpha^2\gamma^2\right)^2+r^2\Delta'\left(\omega^2-\alpha^2\gamma^2\right)^2}
\label{ecm2}
\end{equation}
where $A$ is changed to
\begin{widetext}
\begin{eqnarray}
A'& &=2r\gamma^2\alpha^4-\Delta'\omega^2-\omega\alpha^2\gamma\left(2r\alpha^2-\Delta'\right)\left(L_1+L_2\right)-\alpha^4\left(\gamma^2\Delta'-2r\omega^2\right)L_1L_2-\nonumber\\
& &\frac{1}{2}\sqrt{\frac{r^2\gamma^2\alpha^4-\Delta\omega^2-2\alpha^2\gamma\omega\left(r^2\alpha^2-\Delta\right)L_2-\alpha^4\left(\gamma^2\Delta-r^2\omega^2\right)\dut{L}{2}{2}-r^2\Delta\left(\omega^2-\alpha^2\gamma^2\right)^2}{r^2\gamma^2\alpha^4-\Delta\omega^2-2\alpha^2\gamma\omega\left(r^2\alpha^2-\Delta\right)L_1-\alpha^4\left(\gamma^2\Delta-r^2\omega^2\right)\dut{L}{1}{2}-r^2\Delta\left(\omega^2-\alpha^2\gamma^2\right)^2}}\nonumber\\
& &\left(2r\gamma^2\alpha^4-\Delta'\omega^2-2\alpha^2\gamma\omega\left(2r\alpha^2-\Delta'\right)L_1-\alpha^4\left(\gamma^2\Delta'-2r\omega^2\right)\dut{L}{1}{2}-2r\Delta\left(\omega^2-\alpha^2\gamma^2\right)^2-r^2\Delta'\left(\omega^2-\alpha^2\gamma^2\right)^2\right)\nonumber\\
& &\frac{1}{2}\sqrt{\frac{r^2\gamma^2\alpha^4-\Delta\omega^2-2\alpha^2\gamma\omega\left(r^2\alpha^2-\Delta\right)L_1-\alpha^4\left(\gamma^2\Delta-r^2\omega^2\right)\dut{L}{1}{2}-r^2\Delta\left(\omega^2-\alpha^2\gamma^2\right)^2}{r^2\gamma^2\alpha^4-\Delta\omega^2-2\alpha^2\gamma\omega\left(r^2\alpha^2-\Delta\right)L_2-\alpha^4\left(\gamma^2\Delta-r^2\omega^2\right)\dut{L}{2}{2}-r^2\Delta\left(\omega^2-\alpha^2\gamma^2\right)^2}}\nonumber\\
& &\left(2r\gamma^2\alpha^4-\Delta'\omega^2-2\alpha^2\gamma\omega\left(2r\alpha^2-\Delta'\right)L_2-\alpha^4\left(\gamma^2\Delta'-2r\omega^2\right)\dut{L}{2}{2}-2r\Delta\left(\omega^2-\alpha^2\gamma^2\right)^2-r^2\Delta'\left(\omega^2-\alpha^2\gamma^2\right)^2\right)\nonumber\\
\end{eqnarray}
\end{widetext}
This shows that a singularity in the center-of-mass energy is achieved on the extremal horizon as shown in Fig.(\ref{ecm}) for at most specific values of angular momentum; that is $E_{c.m.}$ is finite for generic values of particle angular momentum. In this way every finite energy value is achieved up to the event horizon and infinite center-of-mass energy is obtained only for some particle collisions on the horizon as in Ref.\cite{p4}.
\newline
Finally a critical angular momentum value is found, namely
\begin{equation}
L_c=\frac{1}{308} \left(253 \sqrt{5}+\frac{\sqrt{55 \left(855 \sqrt[3]{2}+228\,\,2^{2/3}\right)}}{\sqrt[6]{2}}\right)
\end{equation}
The collisional energy $E_{c.m.}$ is plotted in Fig.(\ref{ecm}) where it is shown that infinite $E_{c.m.}$ is obtained only for specific values of angular momentum as expected.
If the angular momentum of both individual particles is greater than the critical angular momentum then they will not reach the horizon at all, and conversely if they both have angular momentum below the critical value then they will fall into the black hole with a finite center-of-mass collisional energy.

\section{III. Conclusion}
In this paper we investigated the effect on the center-of-mass frame energy by colliding two neutral particles of the same mass parameter in a rotating charged cylindrical black hole. In particular we showed how the mechanism found in Ref.\cite{p4}, which was studied in $\mathcal{S}^2$ topology can be transported to an $\mathcal{R}\times\mathcal{S}^1$ topology. Another interesting scenario to investigate would be the toroidal $\mathcal{S}^1\times\mathcal{S}^1$ topology, which may also exhibit this mechanism.
\newline
A particular property of the cylindrical black hole is that the background spacetime is not curved in the axial direction for rotating solutions unlike the spherical case which is curved in every component. Despite this alteration in the calculation the same result follows as is expected, namely a critical angular momentum, $L_c$, is found for extremal charge and rotation, with different values given for the angular momentum of the other particle with only one giving arbitrarily large center-of-mass collisional energy on the horizon while the others do also give this facet however at radii within the horizon. Hence the mechanism in Ref.\cite{p4} is thus shown to be exhibited also for cylindrical black holes in an analogous manner.
\newline
On another note, concerning the $\alpha$ term, in the units used in this paper the condition
\begin{equation}
\alpha>1
\end{equation}
must be satisfied for the arbitrary center-of-mass collisional energy to occur at the horizon. This property emerges by considering the explicit forms of the $E_{c.m.}$ function for different black hole parameters.
\newline
Rotating black holes tend to produce accretion discs about their equator; however, for cylindrical black hole topologies there is no preferred axial value for the equator, and so the likelihood of the energy emission of such a collision is less likely to have interactions with intermediate fields. Using the current toolbox of theoretical particle accelerators, no current design for terrestrial accelerators can produce energies as high as these. It is for this reason that cylindrical black holes could provide the possibility of a high energy physics probe that could explore scales unattainable by current terrestrial accelerators.

\end{document}